%% file: main.tex
\documentclass[sigconf]{acmart}
\pdfoutput = 1

\AtBeginDocument{%
  }

\copyrightyear{2024}
\acmYear{2024}
\setcopyright{rightsretained}
\acmConference[LAMPS '24]{Proceedings of the 1st ACM Workshop on Large AI Systems and Models with Privacy and Safety Analysis}{October 14--18, 2024}{Salt Lake City, UT, USA}
\acmBooktitle{Proceedings of the 1st ACM Workshop on Large AI Systems and Models with Privacy and Safety Analysis (LAMPS '24), October 14--18, 2024, Salt Lake City, UT, USA}
\acmDOI{10.1145/3689217.3690613}
\acmISBN{979-8-4007-1209-8/24/10}

\input{commands-reduce}

\begin{document}

% \title{\tool: An AI-Powered System for Automated Open-Source Cyber Threat Intelligence Gathering and Management}

\title[\tool: An AI-Powered System for Automated Open-Source Cyber Threat Intelligence \\ Gathering and Management]{\tool: An AI-Powered System for Automated Open-Source Cyber Threat Intelligence Gathering and Management}

\author{Peng Gao}
\affiliation{%
  \institution{Virginia Tech}
  \city{Blacksburg}
  \state{VA}
  \country{USA}}
\email{penggao@vt.edu}

\author{Xiaoyuan Liu}
\affiliation{%
  \institution{University of California, Berkeley}
  \city{Berkeley}
  \state{CA}
  \country{USA}}
\email{xiaoyuanliu@berkeley.edu}

\author{Edward Choi}
\affiliation{%
  \institution{University of California, Berkeley}
  \city{Berkeley}
  \state{CA}
  \country{USA}}
\email{edwardc1028@berkeley.edu}

\author{Sibo Ma}
\affiliation{%
  \institution{University of California, Berkeley}
  \city{Berkeley}
  \state{CA}
  \country{USA}}
\email{siboma@berkeley.edu}

\author{Xinyu Yang}
\affiliation{%
  \institution{Virginia Tech}
  \city{Blacksburg}
  \state{VA}
  \country{USA}}
\email{xinyuyang@vt.edu}

\author{Dawn Song}
\affiliation{%
  \institution{University of California, Berkeley}
  \city{Berkeley}
  \state{CA}
  \country{USA}}
\email{dawnsong@berkeley.edu}

\thanks{The first two authors contributed equally to this research.}

\begin{abstract}

Open-source cyber threat intelligence (\cti) has become essential for keeping up with the rapidly changing threat landscape. However, current \cti gathering and management solutions mainly focus on structured Indicators of Compromise (IOC) feeds, which are low-level and isolated, providing only a narrow view of potential threats.
Meanwhile, the extensive and interconnected knowledge found in the unstructured text of numerous \cti reports (\eg security articles, threat reports) available publicly is still largely underexplored.

To bridge the gap, 
we propose \tool, an automated system for \cti gathering and management. \tool efficiently collects a large number of \cti reports from multiple sources, leverages specialized AI-based techniques to extract high-quality knowledge about various threat entities and their relationships, and constructs and continuously updates a threat knowledge graph by integrating new \cti data. 
\tool features a modular and extensible design, allowing for the addition of components to accommodate diverse \cti report structures and knowledge types.
Our extensive evaluations demonstrate \tool's practical effectiveness in enhancing threat knowledge gathering and management.

\end{abstract}




\maketitle

\section{Introduction}
\label{sec:intro}

Despite the dramatic growth in expenses on operational security, we are still witnessing  numerous targeted cyber attacks. These sophisticated attacks leverage various types of exploits and vulnerabilities to penetrate into the system
and steal valuable data. Many high-profile businesses were plagued 
with huge losses~\cite{equifax,target}. 
To counter these attacks, it is crucial to always remain aware of the fast-evolving cyber threat landscape and gain up-to-date knowledge about the dangerous threats. For this reason, security researchers and practitioners actively gather and summarize knowledge about cyber threats from past incidents and share the knowledge to the public.
Providing a form of evidence-based knowledge, such open-source cyber threat intelligence (\cti)~\cite{li2019reading} has the potential to empower various downstream defensive solutions
and has received wide attention.

However, existing \cti gathering and management solutions
are inadequate for the increasing complexity and diversity of cyber threats. They mainly focus on collecting and disseminating \emph{structured} Indicator of Compromise (IOC) feeds~\cite{liao2016acing}.
IOCs are forensic artifacts of intrusions such as hashes of malware samples, names of malicious files/processes, 
and IP addresses and domains of command-and-control (C\&C) servers. Some examples of platforms that share IOCs are PhishTank~\cite{phishtank} and OpenPhish~\cite{openphish} for phishing URLs and Abuse.ch~\cite{abusech} for malware names and hashes. However, these low-level and disconnected IOCs are unable to reveal the complete threat scenario, such as how the threat unfolds into multiple steps, which is common in most sophisticated attacks nowadays~\cite{gao2021enabling}. Defensive solutions that rely on these IOCs are easy to evade when the attacker changes her tools and their signatures~\cite{li2019reading}.

\begin{figure*}[t]
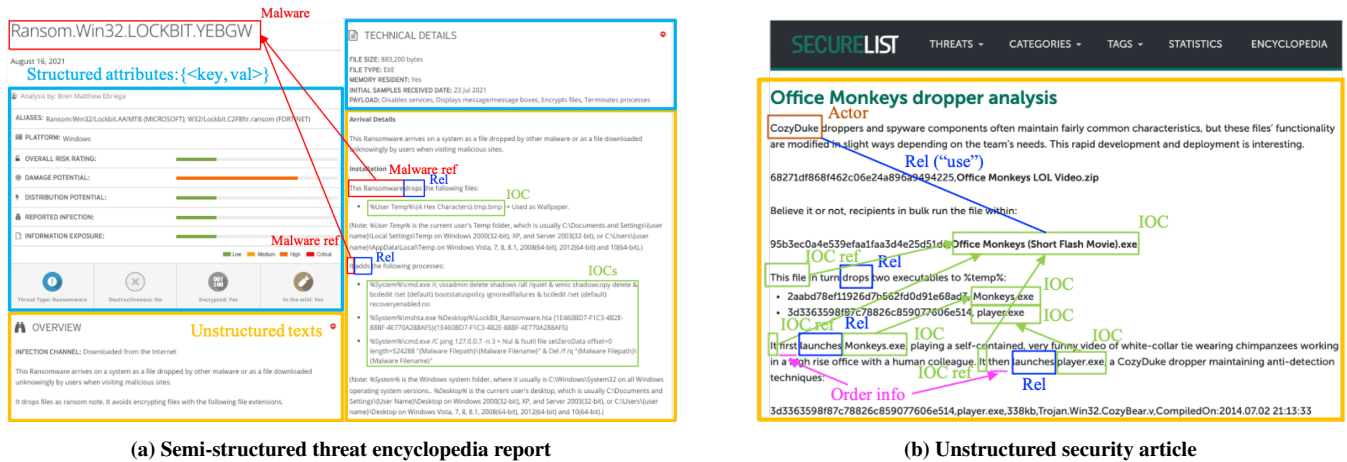

\begin{subfigure}[t]{0.5\textwidth}
  \centering
  \includegraphics[width=\textwidth]{figs/mot\_ex1.png}
    \caption{Semi-structured threat encyclopedia report}
  \label{fig:mot:a}
\end{subfigure}\hfill
\begin{subfigure}[t]{0.44\textwidth}
  \centering
  \includegraphics[width=\textwidth]{figs/mot\_ex2.png}
    \caption{Unstructured security article}
  \label{fig:mot:b}
\end{subfigure}
\caption{Example \cti reports that contain rich threat knowledge. (a) Semi-structured report snippet~\cite{trendmicro-example} from the Trend Micro threat encyclopedia. The report describes the ransomware, \texttt{Ransom.Win32.LOCKBIT.YEBGW}. (b) Unstructured report snippet~\cite{securelist-example} from the Securelist blog. The report describes the \texttt{CozyDuke} threat actor.}
\label{fig:mot}
\vspace{-0ex}
\end{figure*}

In contrast, \emph{\textbf{a large number of unstructured \cti reports have been overlooked}},
These reports are composed and shared by security researchers and practitioners on public websites to summarize threat behaviors in \emph{natural language text}. Some examples of \cti reports are threat encyclopedia pages~\cite{trendmicro-encyclopedia,kaspersky-encyclopedia}, security articles and blogs~\cite{symantec-threat-intel,securelist}, security news~\cite{sophosnews}, etc.
These reports contain not only IOCs, but also other types of \emph{threat knowledge entities}, such as threat actors, adversary tactics, techniques, and procedures (TTPs). Moreover, these reports contain the \emph{semantic relationships between entities} that indicate their interactions (\eg the \ent{read} relationship between two IOCs \ent{/bin/tar} and \ent{/etc/passwd} implies the attacker gathers user credentials into an archive file to be sent out). These relationships provide a form of connected knowledge with more context about cyber threats, which is critical to uncovering multi-step threat behaviors. Studies have shown that defenses based on such connected knowledge (\eg IOC interactions~\cite{gao2021enabling}) are more robust, as they capture the threat behaviors that are aligned to the adversary goals and are harder to change.
However, existing solutions lack the ability to automatically extract such comprehensive knowledge from natural language \cti text.

\begin{figure*}[t]
    \centering
    \includegraphics[width=\linewidth]{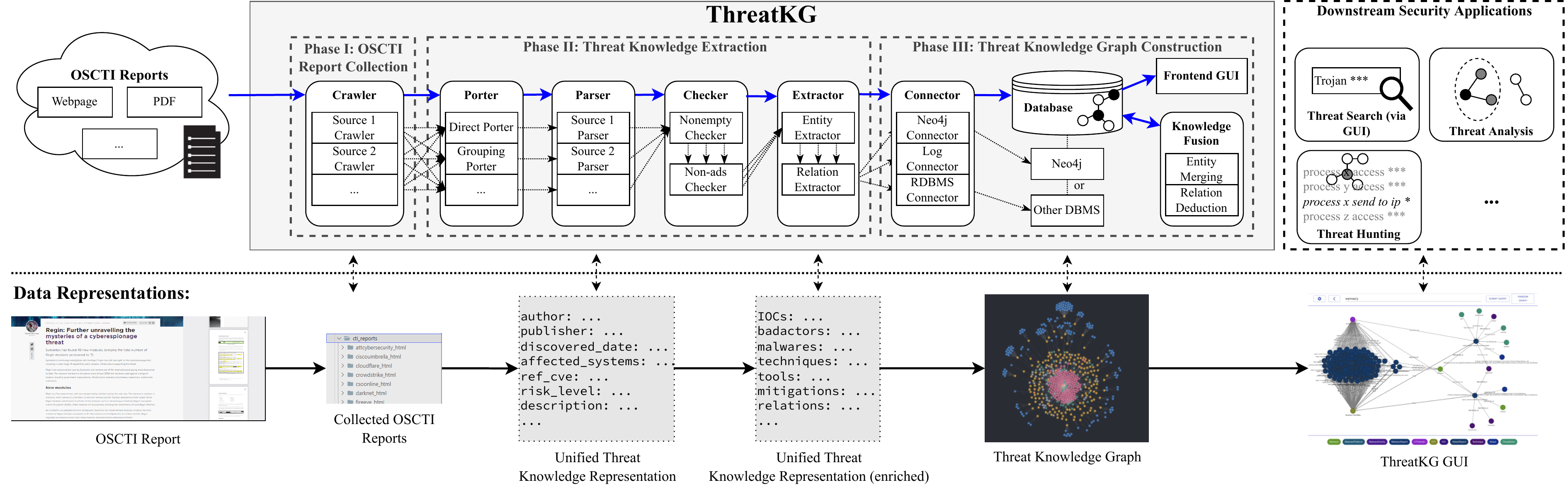}
    \caption{Architecture of \tool. Arrows between system components indicate data flows.}
    \label{fig:architecture}
    \vspace{0ex}
\end{figure*}

\paragraph{Goal and challenges}
We aim to design and build a new system for gathering and managing \cti that can (1) automatically extract high-quality knowledge from a large number of unstructured \cti reports from various sources, and (2) store and organize such knowledge in a unified knowledge base that can provide comprehensive views of different threats. 
The key challenge is three-fold:

(1) \emph{Unified knowledge representation:} To model the threats comprehensively, the system needs to cover a wide range of entity and relation types. Moreover, \cti reports collected from different sources have heterogeneous formats: some reports have structured fields and some reports are mainly composed of text, as shown in \cref{fig:mot}. Also, not all reports from a source are relevant to threats; some of them may be about advertisements or product promotions, as reported in~\cite{liao2016acing}. Therefore, the system needs to handle such diversity, filter out irrelevant reports, and unify the gathered knowledge.

(2) \emph{Accurate knowledge extraction:} Accurately extracting threat knowledge from natural language text is a challenging task. This is because of the presence of many nuances specific to the security context, such as special characters (\eg dots, underscores) in IOCs. These nuances limit the performance of most off-the-shelf natural language processing (NLP) tools for information extraction.
Moreover, collecting large annotated corpora is critical for training knowledge extraction models. A key challenge for the threat knowledge extraction domain is the lack of labeled datasets that encompass the diverse range of entity and relation types that we focus on.

(3) \emph{Efficient knowledge management:} New \cti reports are being released every day that contain fresh threat knowledge. 
The system needs to continuously collect the latest reports from multiple \cti sources, gather new knowledge, and 
integrate the knowledge to update its knowledge base. The system also needs to be extensible to incorporate new \cti sources and report formats.

\paragraph{Contributions}
We propose \tool ($\sim$26K lines of code), a system for automated \cti gathering and management.
\tool collects a large number of \cti reports from various sources, filters out irrelevant reports and extracts high-fidelity threat knowledge using AI-based techniques, constructs a \emph{threat knowledge graph}, and updates the knowledge graph by continuously ingesting new data.

To model the threats comprehensively, \tool employs a hierarchical threat knowledge ontology that covers a wide range of entities. The relations between these entities provide information about both detailed threat behaviors and high-level threat contexts. 
To handle diverse \cti report formats and generalize well to new formats and knowledge, \tool separates the knowledge extraction process into source-dependent parsing and source-independent extraction.
To deeply understand the semantic meaning and connections between targeted entities, \tool employs specialized deep learning-based techniques 
to handle the nuances and accurately extract the knowledge.
We further leverage data programming techniques~\cite{ratner2016data} to programmatically build synthetic annotations to train these models. 
\tool employs an \emph{extensible system architecture} to continuously gather new knowledge in a timely manner. The architecture coordinates all individual components in a modular way, enabling efficient parallelization.
Existing components can be turned off or updated, and new components can be easily added following the common interface. This allows \tool to incorporate new \cti sources or knowledge types. 

To demonstrate the use cases of the threat knowledge graph, we develop two applications: (1) a graphical user interface application for visualizing, exploring, and searching the knowledge graph (demo video~\cite{threatkg-demo}); (2) a question answering system to facilitate threat knowledge acquisition using natural language (demo video~\cite{threatqa-demo}).

\section{System Overview}
\label{sec:overview}

\cref{fig:architecture} illustrates the architecture of \tool, which consists of three phases: (1) \cti report collection, (2) threat knowledge extraction, and (3) threat knowledge graph construction. Each phase consists of one or several processing steps (\eg Parser, Extractor). In Phase I, \tool collects \cti reports from a wide range of sources (Crawler). In Phase II, \tool aggregates multi-page report files (Porter), parses the reports (Parser), filters out non-threat reports (Checker), and extracts threat knowledge (Extractor). In Phase III, \tool constructs a \seckg and stores it in the database.
\tool is fully automated. It collects new reports periodically and incrementally, and extracts new knowledge from them. It then integrates the new knowledge into the threat knowledge graph.

\cref{fig:mot:a} shows a report snippet from the Trend Micro threat encyclopedia~\cite{trendmicro-example} that describes the \ent{Ransom.Win32.LOCKBIT.YEBGW} ransomware.
The report has a semi-structured format, with some structured fields that provide attributes of the malware (\eg aliases and platform) and some natural language text that provides detailed behaviors (\eg dropping a file). 
\cref{fig:mot:b} shows a report snippet from the Securelist blog~\cite{securelist-example} that describes the \ent{Office Monkeys} dropper used by the \ent{CozyDuke} threat actor. It primarily contains natural language text.
We can observe that \cti reports have diverse formats and contain rich threat knowledge. We annotated representative entities and relations in the report snippets. Some entity-relation triplets reveal specific threat behavior steps, such as <\ent{Office Monkeys (Short Flash Movie).exe}, \ent{launch}, \ent{player.exe}>. The text may also indicate the sequential order of some steps, such as ``...first...then...'' in \cref{fig:mot:b}. Some triplets provide high-level threat contexts, such as the \ent{CozyDuke} actor \ent{uses} the \ent{Office Monkeys (Short Flash Movie).exe} dropper file to perform the attack. These relations may not be explicitly expressed by words in the text.
We take care of the extration of the temporal order using dependency parsing and the relations that are not explicitly expressed by words using neural relation extraction.

\section{Report Collection \& Ontology}
\label{sec:ontology}

\subsection{\cti Report Crawlers}
\label{subsec:collection}

We have developed a robust multi-threaded crawler framework that manages crawlers to collect \cti reports from various security websites, given in \cref{table:oscti-sources-and-statistics} (Appendix). These websites include threat encyclopedias~\cite{trendmicro-encyclopedia,kaspersky-encyclopedia}, enterprise security blogs~\cite{symantec-threat-intel,securelist}, influential personal security blogs~\cite{shneieronsecurity}, security news~\cite{sophosnews}, etc. They provide a rich source of threat knowledge, covering different types of threats such as malware, vulnerabilities, and attack campaigns.

Our crawler framework can handle the specific layout structure of each website and collect report URLs for fetching the content. It can deal with both static pages and dynamically generated content (\eg ``View More'' in~\cite{symantec-threat-intel}). 
The framework schedules periodic execution and reboot after failure for each crawler, ensuring robustness and reliability. To improve the crawling efficiency, the framework employs a multi-threaded design that allows parallel execution of multiple crawlers, as well as fetching multiple reports for each crawler.
With \tool's extensible architecture, new \cti sources can be easily added by adding a corresponding crawler and a parser.

To expand the knowledge coverage, we additionally collected \cti reports from APTnotes~\cite{aptnotes}, a repository of publicly-available reports related to malicious campaigns/activities/software that have been associated with vendor-defined APT groups. These reports are in PDF format and are typically longer and more detailed than the security webpages, which provide complementary threat knowledge. 

These \cti sources provide different kinds of threat knowledge, which we categorize into three broad types: malware reports, vulnerability reports, and attack reports.

\begin{itemize}[itemsep=1pt, topsep=1pt, partopsep=1pt, listparindent=\parindent, leftmargin=*]

\item \emph{Malware reports} and \emph{vulnerability reports} are semi-structured reports that contain knowledge about malware or vulnerabilities. They are collected from threat encyclopedias, such as~\cite{trendmicro-encyclopedia,kaspersky-encyclopedia}. These reports usually have a title indicating the name of the malware/vulnerability entity, followed by structured fields indicating the attributes of the entity and a natural language description of the behaviors of the entity. \cref{fig:mot:a} shows an example malware report snippet on the \ent{Ransom.Win32.LOCKBIT.YEBGW} ransomware.

\item \emph{Attack reports} are unstructured reports that contain knowledge about attack campaigns. They are collected from security blogs and news, such as~\cite{symantec-threat-intel,securelist,shneieronsecurity,sophosnews}. These reports mainly contain natural language text describing the context and behaviors of attack campaigns. \cref{fig:mot:b} shows an example attack report snippet on the \ent{CozyDuke} APT attack. \ent{CozyDuke} is the name of the threat actor/group that is responsible for the attack.

\end{itemize}

\subsection{Hierarchical Threat Knowledge Ontology}
\label{subsec:ontology}

\begin{figure}[t]
    \centering
    \includegraphics[width=.8\linewidth]{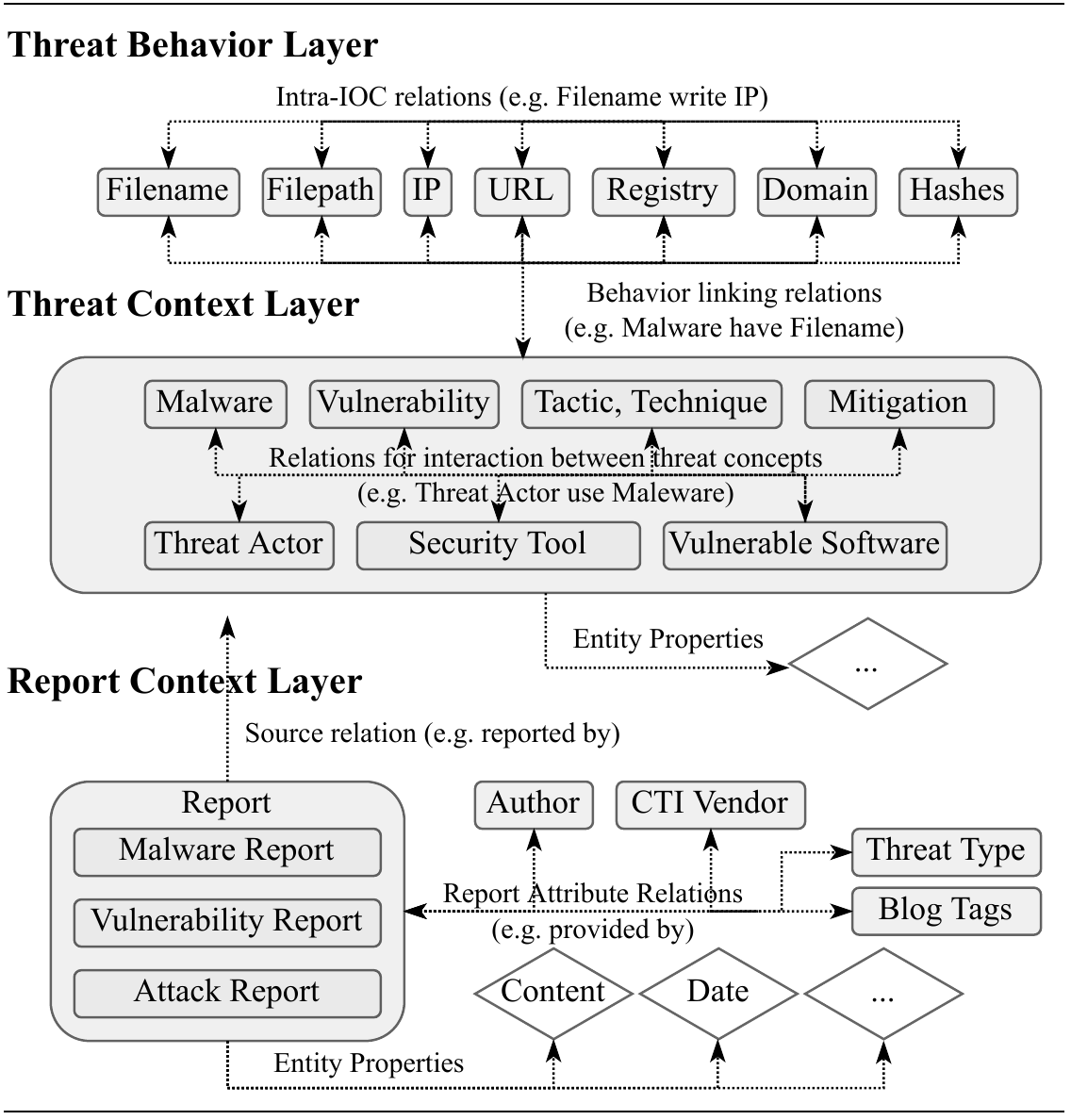}
    \caption{Hierarchical threat knowledge ontology}
    \label{fig:ontology}
    \vspace{-2ex}
\end{figure}

To model the threats comprehensively, we enumerate key knowledge pieces in our collected reports and design a hierarchical threat knowledge ontology. Our ontology (shown in \cref{fig:ontology}) consists of three layers and covers various entities and relations for both low-level threat behaviors and high-level threat contexts.

The report context layer contains report-level knowledge. We create an entity 
for each report and associate it with attributes such as title, URL, publication date, etc. These entities help threat analysts connect other entities (\eg malware, IOCs, TTPs) from the same report and form a comprehensive view of the threat. Threat analysts can also follow the URL attribute to view the original report and obtain more context. Moreover, we create entities for the specific authors and CTI vendors who write and create the reports.

The threat behavior layer contains knowledge about low-level threat behaviors, which are represented by IOCs and their relations. Previous studies~\cite{satvat2021extractor,gao2021enabling} have shown that these relations reveal how the threat progresses through connected steps. This knowledge can help identify system call events (\eg process reading a file) that belong to the attack sequence.
For example, in \cref{fig:mot:b}, two filename IOCs, \ent{Office Monkeys (Short Flash Movie).exe} and \ent{player.exe}, have a \ent{launch} relation. Therefore, we consider various types of IOCs and their interaction verbs (\eg read, write, open, send) as their relations in the threat behavior layer. Example IOC types are filename, filepath, IP, URL, domain, registry, and MD5/SHA1/SHA256 hashes.

The threat context layer provides high-level contexts essential for a comprehensive understanding. This layer includes various entities, including:
(1) malware,
(2) vulnerabilities, 
(3) threat actors (\eg \ent{CozyDuke} APT actor~\cite{securelist-example}),
(4) tactics and techniques (\eg \ent{spearphishing link}~\cite{mitre-attack}),
(5) vulnerable software (\eg \ent{Microsoft Word}),
(6) security-related tools (\eg \ent{Mimikatz}), and
(7) mitigations (\eg \ent{data backup}),
These entities have different types of relations among them. We use \ent{TYPE_ENT} to denote an entity placeholder of the type ``TYPE''. Some examples of entity-relation triplets in this layer are <\ent{ACTOR_ENT}, \ent{use}, \ent{MALWARE_ENT}>, <\ent{ACTOR_ENT}, \ent{use}, \ent{TOOL_ENT}>, and <\ent{SOFTWARE_ENT}, \ent{has}, \ent{VULNERABILITY_ENT}>.

Entities in different layers can be related. For example, entities in the threat behavior layer and the threat context layer that are extracted from the same report are related to the report entity through a \ent{reported_in} relation. In \cref{fig:mot:a}, the malware entity \ent{Ransom.Win32.LOCKBIT.YEBGW} is related to several filepath IOC entities through an \ent{add} relation (after we perform coreference resolution). In \cref{fig:mot:b}, the threat actor entity \ent{CozyDuke} is related to the filename IOC entity \ent{Office Monkeys (Short Flash Movie).exe} through a \ent{use} relation. Entities can also have attributes in the form of key-value pairs (\eg type of a malware, version of a vulnerable software). The three layers of ontology collectively model the threats from multiple dimensions and in different granularities.

%%%%%%%%%%%%%%%%%%%%%%%%%%%%%%%%%%
\section{Threat Knowledge Extraction}
\label{sec:extraction}

\subsection{Report Parsing and Relevance Checking}
\label{subsec:parser}

The crawlers collect the \cti reports and the porters aggregate them into multi-page files. 
Each \cti source has a specific structure, so we use different parsers to parse them (\ie parsers are source-dependent). The parsers convert each report into a unified threat knowledge representation (\tkr), which is a JSON schema. We create this schema by iterating through \cti reports and adding fields for new types of knowledge.
It has fields such as title, author, and placeholders for entities and relations. The parsers also parse the unstructured text blocks and put them into the \tkrs. The extractors then enrich the \tkrs by extracting more entities and relations from the unstructured text. Having a unified representation increases the system's modularity and extensibility; new components can be easily added as long as they work with the same schema design.

There could be irrelevant reports that do not contribute knowledge to model cyber threats, such as empty pages, advertisements, and product promotions.
To filter out these reports, \tool employs a set of checkers that operate on the \tkrs produced by the parsers. 
Empty web pages can be easily filtered out. For ads and other irrelevant reports, we model the relevance checking process as a binary classification task and construct AI-based checkers.
We engineer a set of useful features, including:
(1) Keyword count and density: We count the number and proportion of keywords in the report title and body. We use a list of keywords from MITRE ATT\&CK~\cite{mitre-attack}, such as threat actors, malware, tools, techniques, etc.;
(2) IOC count and density: We extract IOCs using regex rules~\cite{ioc-parser}. We only consider the report body, as most of the titles do not contain IOCs;
(3) Report length: We measure the number of words in the report. We observe that a longer report is more likely to contain threat behaviors;
(4) TF-IDF values: We calculate the TF-IDF (term frequency–inverse document frequency) value for each token in the report to prioritize frequent, unique tokens.
We train various machine learning models (\eg SVM, Random Forest, XGBoost, LightGBM) using these features and compare their performance in \cref{subsec:eval-rq1}.

\subsection{Threat Entity Extraction}
\label{subsec:ner}

The extractors are source-independent; every extractor extracts the targeted knowledge from the text in all reports, and the extraction does not depend on the specific layout structure of each source. By decoupling the knowledge extraction process into source-dependent parsing and source-independent extraction, \tool can easily incorporate new \cti sources (via adding crawlers and parsers) and new knowledge entities and relations (via adding extractors).

For IOCs, we construct a set of regex rules~\cite{ioc-parser} that cover a wide range of IOC types.
\tool incorporates these rules in a rule-based IOC extractor.
For other types of entities
that are hard to define using rules, \tool employs a deep learning-based extractor to perform neural named entity recognition (NER). NER is an information extraction task that aims to identify and categorize named entities in text into pre-defined classes. Deep learning-based approaches have an advantage over conventional methods like Hidden Markov Model,
as they do not require manual feature engineering and can better capture the semantic meaning and hidden patterns of text, resulting in more accurate extraction.

Compared to general text, \cti text has many nuances that are specific to the security context, such as dots, underscores, spaces, and slashes in IOCs. These nuances can cause errors in most basic NLP modules (\eg sentence segmentation, tokenization), and affect the extraction techniques that rely on these modules. To deal with these nuances, we substitute the IOCs with meaningful words that fit the natural language context (\eg word ``FILE’’ for a file IOC token) before applying neural NER, and replace them with the original IOCs after extracting other entities. 

To perform neural NER on \cti text, we construct a Bidirectional LSTM-CRF (BiLSTM-CRF) model~\cite{lample2016neural} for our deep learning-based entity extractor.
(1) First, we tokenize each input sentence and convert each token into an embedding vector using one-hot encoding.
(2) Then, we feed the embeddings to the bidirectional LSTM (BiLSTM) layer, which has two LSTM networks that process the input sentence from both directions. LSTM (Long-Short Term Memory)~\cite{hochreiter1997long} is known for its capability to capture long-range dependencies of tokens. However, a single LSTM can only access information from the past context. For tasks like NER, it is important to understand the context of a token from both past and future contexts. Therefore, we use another LSTM, which processes the information in a reverse order. The BiLSTM acts as a deep feature extractor that captures the sequential relationships among the input tokens.
(3) The outputs from the BiLSTM are passed to a linear layer, which maps the features extracted by the BiLSTM from the feature space to the tag space. After mapping, the outputs are passed to a Conditional Random Field (CRF) layer, which predicts the optimal, joint tags for the whole sentence.

To prepare the training corpus, we use the BIO format to label tokens with entity tags. The BIO format uses three types of prefixes: (i) B- prefix, for a token at the start of an entity chunk, (ii) I- prefix, for a token within a chunk, and (iii) O- prefix, for a token outside a chunk. Some examples of tags are B-BADACTOR (for threat actor), B-MALWARE, B-TOOL, B-TECHNIQUE, B-MITIGATION, etc.

\subsection{Threat Relation Extraction}
\label{subsec:re}

\paragraph{Dependency parsing-based relation extraction}
As discussed in \cref{sec:overview}, some relations are directly \emph{associated with verbs} that describe the interaction between two entities (\eg the \textsf{drop} relation between the malware entity and an IOC in \cref{fig:mot:a}, the \textsf{launch} relation between the IOCs \ent{Office Monkeys (Short Flash Movie).exe} and \ent{player.exe} in \cref{fig:mot:b}). 
Some other relations are \emph{not explicitly} indicated by any words in the text (e.g., the \textsf{use} relation between \ent{CozyDuke} and \ent{Office Monkeys (Short Flash Movie).exe} in \cref{fig:mot:b}).

For the first type, we design a dependency parsing-based relation extractor to identify the verbs that express the interaction between two entities.
We adopt dependency parsing~\cite{jurafsky2000speech} to analyze the grammatical structure of a sentence and constructs a dependency tree.
Then, we use a set of dependency grammar rules~\cite{gao2021enabling} to extract the subject-verb-object relations between the extracted entities.
We also extract the \emph{temporal order} of the interaction steps by looking for specific tokens (\eg ``first'', ``then''), if present.

\paragraph{Neural relation extraction}
For the second type, the dependency parsing-based approach will not work, as these relations do not have explicit verbs in the text. Instead, we model the relation extraction as a \emph{multi-class classification task}: given a sentence that contains two entities recognized by our entity extractors, we determine the relation class between them. The entities include the IOCs recognized by our IOC rules and the other entities recognized by our BiLSTM-CRF model. Some examples of relation classes are \ent{USE} (\ie using something to achieve a goal), \textsf{CREATE} (\ie creating or making something that did not exist before), \textsf{BREAK} (\ie stopping or preventing something from happening), \textsf{FIND} (\ie finding or locating something), and \textsf{ALIAS} (\ie two entities being synonyms). 
A complete list of relation classes in given in \cref{table:relation} (Appendix).  
In general, two entities could have a relation when they co-occur within a certain distance. These entities could co-occur in the same sentence or in different sentences. In our current implementation, we focus on entities that co-occur in the same sentence, as they are more likely to produce high-quality relations based on our observations.

To perform neural relation extraction (RE) on \cti text, we construct a Piecewise Convolutional Neural Networks model with selective attention mechanism (PCNN-ATT) for our deep learning-based relation extractor. 
%~\cite{lin-etal-2016-neural} 
The Piecewise Convolutional Neural Networks (PCNN) model~\cite{zeng2015distant} is a variation of the Convolutional Neural Networks (CNN) model that is widely used for image and text classification tasks. 
However, PCNN is specially designed for relation extraction: it splits a sentence into three parts by the two entities and applies piecewise max pooling to each part, instead of using a single max pooling to merge features as in CNN. This way, PCNN can capture the structural information about the sentence and the two entities, and identify the important tokens between them that indicate the relation.
We convert the sentences into embedding vectors using word embeddings from GoogleNews-vectors-negative300~\cite{googlenews}, position embeddings, and part-of-speech embeddings (indicating the roles of the words). 
Moreover, we use an attention laye on top of the PCNN output to make the model focus on the tokens that are more relevant for relation extraction.

Before extracting relations, \tool performs coreference resolution~\cite{lee2017end} to find all the expressions (\eg pronouns) in the text that refer to the same entity. \cref{fig:mot} shows some examples of entity coreferences indicated by the arrows. This way, the relation extractor can use the information from the resolved entities and the extracted triplets can form a comprehensive view of threat knowledge.

\subsection{Data Programming}

To train deep learning-based models for NER and RE, we need a large annotated corpus. However, manually annotating such a corpus is costly: for NER, we need to tag each token in the text with a label in the BIO format; for RE, we need to label each sentence in the text with a relation class and the types and location spans of the entities in the sentence. 
Unlike other information extraction domains that have plenty of labeled datasets, there are no large annotated corpora for the threat knowledge extraction domain. To reduce the cost of obtaining supervision, we leverage data programming~\cite{ratner2016data}, which synthesizes annotations \emph{programmatically} using unsupervised modeling of sources of weak supervision. Specifically, data programming obtains the domain knowledge expressed by subject matter experts through labeling functions (which could be noisy rules based on heuristics), and then denoises and integrates these sources of weak supervision to synthesize annotations.
We use Snorkel~\cite{snorkel-web}, an open-source implementation of data programming, to programmatically create large training sets for our NER and RE tasks. 
Snorkel does not need any labeled data for training (\ie unsupervised): after we construct labeling functions, it automatically learns and assigns weights to the labeling functions and produces a single set of noise-aware confidence-weighted labels for the input samples.

The most important step in synthesizing good annotations is to define noisy but useful labeling functions, which we spent most of our efforts on. To synthesize annotations for the NER task, we create labeling functions based on our curated list of entity keywords. For example, we construct the list of threat actors, malware, techniques, and tools from MITRE ATT\&CK~\cite{mitre-attack}. To synthesize annotations for the RE task, we create labeling functions based on distant supervision and checking the entity types and keywords existence.

Distant supervision~\cite{mintz2009distant} is a technique that uses an existing knowledge base to generate training data. The idea is that if two entities have a fact in the knowledge base, we can label any sentence that contains them as a positive example for the relation that the fact represents. This way, we can create a large number of (noisy) labeled sentences. 
For example, Freebase contains the fact that Barack Obama and Michelle Obama are married. We use this fact to label any sentence that has ``Barack Obama'' and ``Michelle Obama'' as a positive example for our marriage relation. In our threat knowledge extraction task, we use MITRE ATT\&CK, which is a manually curated knowledge base by security experts for cyber adversary behaviors and can be downloaded as a JSON file. For example, for a sentence that has a threat actor entity and a malware entity, if the two entities are in the MITRE ATT\&CK and have the ``use'' relation type, we label the sentence with the \ent{USE} relation class.

We also create labeling functions based on heuristic rules that assign relation labels based on the entity types and the presence of keywords. For example, for the \ent{ALIAS} relation, we verify that the two entities belong to the same type and look for keywords such as ``alias'' or ``aka''. By leveraging data programming, we can generate a large amount of training data with low human effort.

\section{Extensible System Architecture}
\label{sec:backend}

\paragraph{Threat knowledge graph construction}
After the extractors enrich the \tkrs, \tool constructs the threat knowledge graph from them and stores it in the database for persistence. Storing the \tkrs directly is inefficient and makes it hard for end users (\eg threat analysts) to understand and analyze them. 
Therefore, \tool converts these intermediate representations to match the threat knowledge ontology, which has clear and concise semantics for entities and relations. This ontology is designed separately from the \tkrs. 
\tool then integrates the transformed representations into the database using its connectors. 
Currently, \tool uses Neo4j, the leading graph database, as its storage, where nodes are entities and edges are relations. Each node has a category (\eg malware or threat actor), a unique name (\eg specific malware name), and a set of attributes. \tool can easily support new database backends by adding the corresponding connectors without changing the previous components in its processing pipeline.

\paragraph{Modularity and extensibility}
\tool adopts a modular design to make the system extensible, which enables multiple system components in the same processing step to have the same input/output interface. 
For example, \tool uses multiple crawlers to collect \cti reports from various sources, and different porters to import report data from various formats, such as HTML, PDF, and compressed files.
Moreover, \tool supports rich configuration options: the system can be customized through a configuration file, which defines the components to use and the parameters to pass to them (\eg threshold values for NER). With this design, existing components can be easily replaced or added.

We parallelize the system components for the processing steps (\eg crawlers, parsers, checkers, extractors) to improve the system efficiency.
We define the formats of intermediate representations (\ie \tkrs) and make them serializable between different processing steps. These \tkrs are enriched as they pass through the pipeline.

\paragraph{Continuous updating}
\tool is automated and continuously running to provide the latest threat knowledge in a timely manner. The threat knowledge graph is updated incrementally with new reports being collected and new knowledge being extracted and integrated.
Different sources may use different identifiers for the same entity. For example, ``ZQuest'' and ``Z-Quest'' refer to the same adware. To ensure consistency, \tool combines knowledge from multiple sources using knowledge fusion:
\tool scans all the entities and merges facts about the same entity by creating a new entity as the result and moving all relations. 
A key challenge is that entities with similar names may be different. For example, ``Petya'' and ``NotPetya'' are two ransomware with names satisfying a substring relation but are different entities. To address this challenge, \tool uses the contextual information stored with the entity and only merges two entities when they have a similar name (\eg semantic similarity computed using word embeddings) above a threshold, no conflicts in their attributes, and operate in a similar environment (\eg the same platform). By using contextual information and avoiding conflicts, \tool minimizes the information loss in its knowledge fusion, while providing a consistent and comprehensive view of entities from multiple sources.

\section{Downstream Security Applications}
\label{sec:apps}

Various security applications can be built upon the threat knowledge graph to enhance the defenses. In this section, we present two applications that we built to facilitate threat knowledge graph visualization and exploration and threat knowledge acquisition.

\subsection{GUI for Threat Visualization and Exploration}
\label{subsec:ui}

To facilitate threat search and threat knowledge graph exploration, we built a GUI using React and Elasticsearch.
The GUI interacts with the threat knowledge graph stored in the Neo4j database and provides various types of interactivity.
As illustrated in our demo video~\cite{threatkg-demo}, the user can zoom in/out and drag the canvas to adjust the view, click on a node or an edge to see the detailed information, and search information by keywords (using Elasticsearch) or Cypher queries (using Neo4j Cypher engine).
Once the user drags a node, the GUI responds to the node movements to prevent overlap through an automatic graph layout using the Barnes-Hut algorithm~\cite{barnes1986hierarchical}, which calculates the nodes' approximated repulsive force based on their distribution. 
The dragged nodes will lock in place but are still draggable if selected.
This feature helps the user create custom graph layouts for visualization.

The GUI also supports convenient threat knowledge graph navigation. The user can double-click on a node to expand or collapse its neighboring nodes. If the neighboring nodes are not in the view, they will appear when the node is double-clicked. If the neighboring nodes or any downstream nodes are in the view, they will disappear when the node is double-clicked again. The user can also adjust the number of nodes and the maximum number of neighboring nodes displayed for each node, and go back to the previous graphs displayed.
Our GUI is not tied to the specific database backend, and it can easily switch to a different database (\eg RDF store) while providing the same functionalities.

\subsection{Question Answering System}
\label{subsec:qa}

We built a QA system (named \qatool) on top of \seckg to facilitate threat knowledge acquisition. The user can ask a natural language question about the attributes or connections of a threat, and \qatool will return the answer from the \seckg. \qatool can handle: (1) simple questions that ask for an entity's attribute or related entities, such as ``Which CVE ID is exploited by EternalBlue?'', and (2) complex questions that require multi-hop reasoning, such as ``What common techniques are used by DarkVishnya and Chimera?''

The system follows a three-stage pipeline: 

\begin{itemize}[itemsep=1pt, topsep=1pt, partopsep=1pt, listparindent=\parindent, leftmargin=*]

\item \emph{Stage 1: Entity linking:} 
To link the question to the \seckg, \qatool first identifies the entities in the question using the neural NER approach described in \cref{subsec:ner}. Then, for each entity, it searches for the most similar entity in the \seckg among the entities of the same category. 
\qatool links the entity in the question to the entity in the \seckg with the highest similarity.

\item \emph{Stage 2: Question intent mapping:} \qatool uses a Roberta-based intent classifier~\cite{liu2019roberta} to identify the question's intent. Then, it finds the attribute or relation in the \seckg that matches the question’s intent. The attribute or relation helps \qatool to locate a subgraph of the \seckg that contains the answer.

\item \emph{Stage 3: Query synthesis and answer retrieval:}
\qatool adopts a template-based approach to generate Cypher queries for different types of questions. It has carefully designed query templates (one example shown below) that encode the path between the entity in the \seckg and the target answer. It fills in the linked entities into the query template that matches the asked attribute or relation. Then, it executes the query over the \seckg stored in the Neo4j database and retrieves the final answer. This approach ensures that the query is grammatically correct and reliable.

\end{itemize}

\begin{lstlisting}[style=myStyleMain]  
    // Cypher query template 
    % // Identify techniques used by a threat actor, var1
    % // Return: a list of techniques
    MATCH (var1:Actor)-[rel:USE]->(var2:Technique) 
    WHERE ACTOR_NAME IN var1.name 
    RETURN var2.name
\end{lstlisting}

We built a GUI for \qatool using React, as shown in our demo video~\cite{threatqa-demo}. The GUI displays the results of each QA processing stage:
(1) the recognized question intent (``malware\_type''), with the entities in the question and their categories;
(2) the entity linking result (from the entity name ``Downloader.Slime'' in the question to the malware node in the \seckg);
(3) the synthesized Cypher query and the final answer (``trojan house'').
The GUI also allows the user to edit the Cypher query to investigate the malware further.

\section{Evaluation}
\label{sec:eval}

\begin{table}[t]
\centering
\caption{Statistics of our labeled ground-truth \cti dataset} 
\label{table:labeled-stats}
\begin{adjustbox}{width=.8\linewidth}

\resizebox{\columnwidth}{!}{\begin{tabular}{l lr}
\toprule
\textbf{Data Source} & \textbf{Category} & \textbf{\# Reports}\\ 
\midrule
apt\_notes & APT Reports & 15 \\ 
kaspersky\_threat & Threat Encyclopedia & 45 \\ 
symantec\_threat & Threat Encyclopedia & 45 \\
attcybersecurity & Enterprise Security Blog & 12 \\ 
crowdstrike & Enterprise Security Blog & 6 \\
securelist & Enterprise Security Blog & 7 \\
symantecthreatintelligence & Enterprise Security Blog & 11\\
\hline
\textbf{Total:} & & 141  \\ 
\bottomrule
\end{tabular}}
\end{adjustbox}
\vspace{-1ex}
\end{table}

\begin{table*}[t]
\centering
\caption{Report checker performance (averaged for different classifiers)} 
\label{table:checker-averaged-results}
\begin{adjustbox}{width=.84\linewidth}
\resizebox{\columnwidth}{!}{\begin{tabular}{l|cccc|cccc|cccc}
\toprule
 \multicolumn{1}{c}{} 
 & \multicolumn{4}{c}{\textbf{Symantec Threat Intelligence}}  & \multicolumn{4}{c}{\textbf{Securelist}}  & \multicolumn{4}{c}{\textbf{Webroot}} \\ \hline
\textbf{Training Procedure} & \textbf{Accuracy} & \textbf{F1} & \textbf{FPR} & \textbf{FNR} & \textbf{Accuracy} & \textbf{F1} & \textbf{FPR} & \textbf{FNR} & \textbf{Accuracy} & \textbf{F1} & \textbf{FPR} & \textbf{FNR} \\
\midrule
Source-specific & 93.33\% & 95.38\% & 21.21\% & 0.00\% & 81.14\% & 87.67\% & 53.62\% & 3.77\% & 78.33\% & 86.02\% & 64.10\% & 1.23\% \\ 
Universal & 94.29\% & 95.92\% & 13.64\% & 2.08\% & 80.04\% & 86.99\% & 54.35\% & 5.03\% & 73.75\% & 83.50\% & 76.92\% & 1.85\% \\ 
\bottomrule
\end{tabular}}
\end{adjustbox}
\vspace{-1ex}
\end{table*}

\begin{table*}[t]
\centering
\caption{Source-specific checker results} 
\label{table:checker-source-specific-results}
\begin{adjustbox}{width=0.85\linewidth}
\resizebox{\columnwidth}{!}{\begin{tabular}{l|cccc|cccc|cccc}
\toprule
 \multicolumn{1}{c}{} 
 & \multicolumn{4}{c}{\textbf{Symantec Threat Intelligence}}  & \multicolumn{4}{c}{\textbf{Securelist}}  & \multicolumn{4}{c}{\textbf{Webroot}} \\ \hline
\textbf{Models} & \textbf{Accuracy} & \textbf{F1} & \textbf{FPR} & \textbf{FNR} & \textbf{Accuracy} & \textbf{F1}& \textbf{FPR} & \textbf{FNR} & \textbf{Accuracy} & \textbf{F1} & \textbf{FPR} & \textbf{FNR} \\ 
\midrule
Logistic Regression & 94.29\% & 96.00\% & 18.18\% & 0.00\% & 80.26\% & 87.18\% & 56.52\% & 3.77\% & 80.00\% & 87.10\% & 61.54\% & 0.00\% \\
Random Forest & 94.29\% & 96.00\% & 18.18\% & 0.00\% & 81.58\% & 88.33\% & 60.87\% & 0.00\% & 77.50\% & 85.71\% & 69.23\% & 0.00\% \\
Linear SVM & 94.29\% & 96.00\% & 18.18\% & 0.00\% & 80.26\% & 87.18\% & 56.52\% & 3.77\% & 80.00\% & 87.10\% & 61.54\% & 0.00\% \\
Kernel SVM & 88.57\% & 92.31\% & 36.36\% & 0.00\% & 82.89\% & 88.89\% & 52.17\% & 1.89\% & 80.00\% & 87.10\% & 61.54\% & 0.00\% \\
LightGBM & 94.29\% & 96.00\% & 18.18\% & 0.00\% & 82.89\% & 88.70\% & 47.83\% & 3.77\% & 77.50\% & 85.25\% & 61.54\% & 3.70\% \\
XGBoost & 94.29\% & 96.00\% & 18.18\% & 0.00\% & 78.95\% & 85.71\% & 47.83\% & 9.43\% & 75.00\% & 83.87\% & 69.23\% & 3.70\% \\
\textbf{Average} & 93.33\% & 95.38\% & 21.21\% & 0.00\% & 81.14\% & 87.67\% & 53.62\% & 3.77\% & 78.33\% & 86.02\% & 64.10\% & 1.23\% \\ 
\bottomrule
\end{tabular}}
\end{adjustbox}

\end{table*}

\begin{table*}[t]
\centering
\caption{Universal checker results} 
\label{table:checker-universal-results}
\begin{adjustbox}{width=0.85\linewidth}
\resizebox{\columnwidth}{!}{\begin{tabular}{l|cccc|cccc|cccc}
\toprule
 \multicolumn{1}{c}{} 
 & \multicolumn{4}{c}{\textbf{Symantec Threat Intelligence}}  & \multicolumn{4}{c}{\textbf{Securelist}}  & \multicolumn{4}{c}{\textbf{Webroot}} \\ \hline
\textbf{Models} & \textbf{Accuracy} & \textbf{F1} & \textbf{FPR} & \textbf{FNR} & \textbf{Accuracy} & \textbf{F1} & \textbf{FPR} & \textbf{FNR} & \textbf{Accuracy} & \textbf{F1} & \textbf{FPR} & \textbf{FNR} \\
\midrule
Logistic Regression & 94.29\% & 95.83\% & 9.09\% & 4.17\% & 84.21\% & 89.09\% & 34.78\% & 7.55\% & 82.50\% & 88.52\% & 53.85\% & 0.00\% \\
Random Forest & 94.29\% & 96.00\% & 18.18\% & 0.00\% & 76.32\% & 85.48\% & 78.26\% & 0.00\% & 70.00\% & 81.82\% & 92.31\% & 0.00\% \\
Linear SVM & 97.14\% & 97.96\% & 9.09\% & 0.00\% & 85.53\% & 90.43\% & 43.48\% & 1.89\% & 72.50\% & 83.08\% & 84.62\% & 0.00\% \\
Kernel SVM & 97.14\% & 97.96\% & 9.09\% & 0.00\% & 72.37\% & 83.20\% & 86.96\% & 1.89\% & 75.00\% & 84.37\% & 76.92\% & 0.00\% \\
LightGBM & 91.43\% & 93.88\% & 18.18\% & 4.17\% & 82.89\% & 88.29\% & 39.13\% & 7.55\% & 70.00\% & 81.25\% & 84.62\% & 3.70\% \\
XGBoost & 91.43\% & 93.88\% & 18.18\% & 4.17\% & 78.95\% & 85.45\% & 43.48\% & 11.32\% & 72.50\% & 81.97\% & 69.23\% & 7.41\% \\
\textbf{Average} & 94.29\% & 95.92\% & 13.64\% & 2.08\% & 80.04\% & 86.99\% & 54.35\% & 5.03\% & 73.75\% & 83.50\% & 76.92\% & 1.85\% \\ 
\bottomrule
\end{tabular}}
\end{adjustbox}

\end{table*}

We built \tool ($\sim$26K lines of code) upon several tools: Python for the system architecture, BeautifulSoup and Selenium for the crawlers, scikit-learn and Ray Tune (for hyperparameter optimization) for the checkers, PyTorch for the extractors, Snorkel for data programming, and Neo4j for the storage backend.
We evaluate our system on several aspects, such as the accuracy of the knowledge extraction and the performance of the system. We aim to answer the following key research questions:

\begin{enumerate}[label={\textbf{(RQ\arabic*)}}, leftmargin=*,itemsep=4pt, topsep=5pt]

    \item How well can \tool identify and filter out \cti reports that do not contain any cyber threat information?

    \item  How effectively can \tool extract threat knowledge from the \cti text? How much does the data programming technique enhance the extraction performance?
        
    \item How does \tool compare with other baselines in extracting various types of threat knowledge?

    \item For the runtime performance, is \tool efficient enough to be practical for a real-world deployment?
    
\end{enumerate}

\paragraph{Evaluation setup}
We deployed \tool on a server with an AMD EPYC 7282 CPU (2.80GHz), an Nvidia GRID T4-16Q GPU with 16GB RAM, and Ubuntu 20.04 as the operating system.
To evaluate the accuracy of \tool in extracting threat knowledge from \cti reports, we created a ground-truth labeled dataset from seven different \cti sources, namely: APTnotes attack reports, two threat encyclopedias, and four enterprise security blogs. These sources provide diverse and comprehensive \cti reports that cover various types of threat knowledge. 
We manually labeled $141$ reports from these sources according to the ontology we defined.
\cref{table:labeled-stats} summarizes the statistics of our ground-truth \cti dataset.
For the entities, we used the BIO tagging scheme to mark their boundaries and types.
For the relations, we labeled both the relation verbs (if any) and the relation classes between the entity pairs. 
We have 17 relation classes in total (shown in \cref{table:relation} in Appendix), covering major types of threat behaviors.
Two of our authors performed the labeling task independently and then cross-validated each other’s results and resolved any disagreements.

%%%%%%%%%%%%%%%%%%%%%%%%%%%%%
\subsection{RQ1: Accuracy of Irrelevant Report Filtering}
\label{subsec:eval-rq1}

To evaluate the checker performance, 
we created a dataset by randomly selecting 755 reports from three different \cti sources: Securelist~\cite{securelist}, Symantec Threat Intelligence~\cite{symantec-threat-intel}, and Webroot~\cite{webroot}. Out of these, 517 reports are relevant to cyber threats and 238 reports are irrelevant to cyber threats (\eg reports about advertisements, security products, cybersecurity education).

\cti reports collected from different sources vary in their structures, writing styles, and topics. We wanted to examine how the distributional shift in the training data affects the performance of a classifier that predicts the relevance of a report to cyber threats. We conducted two experiments.
In the first experiment, we trained a source-specific classifier for each source using only the data from that source. We then evaluated the classifier on the same source data.
In the second experiment, we trained a universal classifier using the data from all sources combined. We then evaluated the classifier on each source data individually. 
We used six machine learning classifiers for both experiments: Logistic Regression, Random Forest, Linear SVM, SVM with RBF Kernel, XGBoost, and LightGBM. We split the data for each source into 70-10-20 for train/dev/test sets. We merged the train/dev/test sets from all sources to form the corresponding sets for the universal classifier.

\cref{table:checker-averaged-results} shows the results averaged for different models. \cref{table:checker-source-specific-results,table:checker-universal-results} contain the results for each individual model. We observe:
(1) For source-specific classifiers, the average F1 scores are above 86\% and the average false negative rates (FNRs) are below 3.77\%. The false positive rates (FPRs) are higher. In our problem setting, a high FPR is acceptable as long as the FNR can be sufficiently low, because a high FNR means that many relevant reports (and the contained threat knowledge) are filtered out, while a high FPR just means that the system is conservative in filtering the reports. Note that our goal is to extract as much information as possible without overlooking threat-related articles.
(2) The performance of the universal classifier does not benefit from more training data, and is worse than the source-specific classifiers for some sources (\eg Securelist and Webroot). This verifies the distributional shift problem in different \cti sources that we conjectured previously. Based on these observations, we recommend training classifiers for different sources separately to get better checker performance.

\subsection{RQ2: Accuracy of Knowledge Extraction}
\label{subsec:eval-rq2}

Numerous works have successfully used BiLSTM-CRF to extract entities from natural language text, achieving excellent performance. In this RQ, we primarily focus on evaluating the performance of our relation extractor and the impact of data programming on this task.

Our labeled dataset shown in \cref{table:labeled-stats} contains 7308 relations.
We randomly picked 16 reports and constructed the \emph{test set} using the relations in them. 
In the remaining 125 reports, there are 1219 ``non-other'' relations and 4615 \ent{OTHER} relations (assigned when none of the other relation types match).
This dataset is imbalanced and will negatively impact the trained model performance.
Thus, we under-sampled the \ent{OTHER} relations to make the dataset more balanced. 
After under-sampling, there are 1732 \ent{OTHER} relations left. 
To expand the dataset, we manually labeled 805 additional ``non-other'' relations chosen from the same seven \cti sources.
Finally, we created train/dev split of 87.5\% and 12.5\% respectively from the $2024$ ``non-other'' relations and $1732$ \ent{OTHER} relations.
The aggregated results for all relation classes are presented in \cref{table:re-2-results}. The model achieved an F1 score of $79\%$, which is reasonable considering the difficulty of the multi-class classification task (17 classes) and the limited size of the dataset (i.e., $2024$ ``non-other'' relations that we labeled).

\begin{table}[t]
\centering
\caption{Relation extraction performance (aggregated)} 
\label{table:re-2-results}
\begin{adjustbox}{width=.4\textwidth}

\resizebox{\columnwidth}{!}{\begin{tabular}{l rrrr}
\toprule
& \textbf{Precision} & \textbf{Recall} & \textbf{Accuracy} &\textbf{F1}\\ 
\midrule
\textbf{W/O Data Programming} & 80\% & 78\% & 78\% & 79\% \\
\textbf{W/ Data Programming} & 85\% & 85\% & 85\% & 85\% \\
\bottomrule
\end{tabular}}
\end{adjustbox}

\vspace{-1ex}
\end{table}

\paragraph{Effectiveness of data programming}
To address the data scarcity issue, we applied data programming to generate more training instances. We labeled 2049 additional ``non-other'' relations and used all the 4615 \ent{OTHER} relations from the 125 reports. We split the resulting dataset into train/dev subsets with a ratio of 87.5\% and 12.5\%, respectively. We used the same test set as before, which was manually labeled from 16 randomly selected reports. For correct evaluation setup, the test set only includes manual labels and does not include labels obtained via data programming.

The results in \cref{table:re-2-results} show that data programming significantly improved the relation extraction performance, \emph{from 79\% F1 to 85\% F1}. Moreover, the model performed better on the relation types that had fewer training instances in the previous experiment. For instance, the relation \textsf{INJECT} had an F1 score of 55\% with 222 training instances in the previous experiment, but it increased to 72\% with 558 training instances after data programming. These results confirm the effectiveness of data programming in creating more training data to enhance the model.

\begin{table*}[t]
\centering
\caption{Relation extraction performance (-\#false negatives, +\#false positives) of TTPDrill, EXTRACTOR, and \tool}

\label{table:comparison}
\begin{adjustbox}{width=.85\linewidth}
\resizebox{\columnwidth}{!}{\begin{tabular}{l|ccccc}
\toprule

\multirow{1}{*}{\textbf{CTI reports}} 
% & \multicolumn{4}{c}{\textbf{Entities}} & \multicolumn{3}{c}{\textbf{Relations}} \\

& \multicolumn{1}{l}{\textbf{\# Words}} & \textbf{\# Manually Labeled Relations} & \textbf{TTPDrill} & \textbf{EXTRACTOR} & \multicolumn{1}{l}{\textbf{\tool}} \\
\midrule

hunting-for-linux-library-injection-with-osquery  
& 1431  & 145   & (-105, +341)  & (-128, +207)  & (-47, +47)  \\
android-backdoor-disguised-as-a-kaspersky-mobile-security-app-65534
& 233   &  34   & (-28, +130)  & (-28, +43) & (-16, +16)  \\
peer-peer-poisoning-attack-against-kelihosc-botnet      
& 829     &  43  & (-38, +214)   & (-31, +116)  & (-6, +6)  \\
dragonfly-energy-companies-sabotage 
& 1095    & 104 & (-88, +619)   & (-84, +221)  & (-22, +22)   \\
android-apps-coronavirus-covid19-malicious
& 388    & 30 & (-24, +105)  & (-21, +59)  & (-7, +7)  \\
new-versions-of-the-iexplorer-zeroday-emerge-targeting-defence-and-industri                 
& 317    & 40 & (-33, +99)  & (-30, +55)  & (-15, +15)  \\
Trojan.Win64.Shelma                      
& 17     & 2 & (-0, +13)   & (-1, +1)  & (-0, +0)  \\
analyzing-the-security-of-ebpf-maps      
& 1066     & 105 & (-89, +533)  & (-86, +217)  & (-31, +31)   \\
security-vpn-ios-macos                  
& 400     & 12 & (-9, +187)   & (-8, +112)  & (-2, +2)  \\
duqu-next-stuxnet  
& 502     & 49 & (-44, +266)  & (-38, +89)  & (-18, +18)   \\
Trojan-DDoS.Win32.Nesmed                   
& 12     & 4 & (-4, +6)   & (-3, +1)  & (-2, +2)  \\
inside-geinimi-android-trojan-chapter-one-encrypted-data-and-communication    
& 655     & 11 & (-11, +31)  & (-10, +56)  & (-1, +1)   \\
beapy-cryptojacking-worm-china                   
& 1271     & 123 & (-104, +508)   & (-102, +316)  & (-34, +34)  \\
a-few-words-about-the-hlux-botnet-29806    
& 51     & 11  & (-11, +41)  & (-5, +11)  & (-2, +2)   \\
google-cloud-platform-security-monitoring-with-usm-anywhere                  
& 542     & 47 & (-37, +289)   & (-33, +129)  & (-11, +11)  \\
Alienvault_Scanbox   
& 400    & 28  & (-22, +180)  & (-16, +85)  & (-6, +6)   \\

%%% Precision Recall F1 Precision Recall F1 Precision Recall F1
\midrule
\textbf{Overall Precision}  &  &   & \textbf{0.038} & \textbf{0.198} & \textbf{0.85}  \\
\textbf{Overall Recall} &  & & \textbf{0.199} & \textbf{0.305} & \textbf{0.85} \\
\textbf{Overall F-1 Score} &  & &  \textbf{0.063} & \textbf{0.217} & \textbf{0.85} \\
\bottomrule
\end{tabular}}
\end{adjustbox}
\end{table*}

%%%%%%%%%%%%%%%%%%%%%%%%%%%%%%%%%%%%
%%%%%%%%%%%%%%%%%%%%%%%%%%%%%%%%%%%%
\subsection{RQ3: Comparison with Existing Security Information Extraction Approaches}
\label{subsec:eval-rq3}

We compared \tool with two state-of-the-art security information extraction approaches, TTPDrill~\cite{husari2017ttpdrill} and EXTRACTOR~\cite{satvat2021extractor}., to further evaluate \tool’s effectiveness in extracting threat knowledge. 
We used the same 16 reports that we used for the test set in \cref{subsec:eval-rq2}, which covered a wide range of threat scenarios, such as major OS platforms (Linux, Windows, IOS, and Android), well-known APT campaigns (Stuxnet and Beapy), and common types of cyber threats (malware and cryptojacking attacks). 
We ran TTPDrill and EXTRACTOR on these reports, and used the same \tool model that we trained in \cref{subsec:eval-rq2} for performance comparison.

Note that TTPDrill and EXTRACTOR are more limited than \tool in the scope of threat knowledge extraction.
TTPDrill only extracts threat actions and maps them to TTP categories, while EXTRACTOR only extracts subject-predicate-object triplets that involve IOCs. 
\tool, on the other hand, can extract a wider range of knowledge types, such as various entities (IOCs, threat actors, malware, etc.), relations, and entity attributes, from various \cti sources.
Another important difference between TTPDrill and EXTRACTOR and \tool is that TTPDrill and EXTRACTOR only extract limited knowledge from a \emph{single} \cti report, while \tool can extract threat knowledge from a large number of \cti reports from various sources and construct a threat knowledge graph, through an automated system.
This makes \tool much more comprehensive in capturing the threat landscape

As the types of entities covered by TTPDrill and EXTRACTOR are very limited, in our evaluations, we only compare the relations extracted by TTPDrill and EXTRACTOR with \tool. The results are shown in \cref{table:comparison}.
We can observe that:
(1) EXTRACTOR has a lower performance than \tool in relation extraction, because EXTRACTOR can only extract relations between IOCs.
(2) TTPDrill suffers from a low precision, because it aims to extract threat actions and map them to TTP categories, so it extracts many phrases that may not be relevant to the relation types.

\subsection{RQ4: Runtime Performance}
\label{subsec:eval-rq4}

\begin{table}[t]
\centering
\caption{Runtime performance breakdown}
\label{tab:breakdown}
\begin{adjustbox}{width=.85\linewidth}
\begin{tabular}{l|r|r|lr}
\toprule
\textbf{Stage} &
  \multicolumn{1}{l|}{\begin{tabular}[c]{@{}l@{}}\textbf{Total} \\ \textbf{Processing} \\ \textbf{Time (h)}\end{tabular}} &
  \multicolumn{3}{l}{\textbf{Percentage}} \\ \midrule
Porter & 0.54 & \multicolumn{3}{r}{0.6\%}                                                                     \\ \hline
Checker  & 0.03 & \multicolumn{3}{r}{0.0\%}                                                                     \\ \hline
Parser  & 1.45 & \multicolumn{3}{r}{1.7\%}                                                                     \\ \hline
\multirow{5}{*}{Extractor} &
  \multirow{5}{*}{85.26} &
  \multirow{5}{*}{97.7\%} &
  \begin{tabular}[c]{@{}l@{}}Content relevance \\ analysis\end{tabular} &
  2.1\% \\ \cline{4-5} 
       &      &  & \begin{tabular}[c]{@{}l@{}}Dependency parsing for \\ IOC relation extraction\end{tabular} & 83.1\% \\ \cline{4-5} 
       &      &  & \begin{tabular}[c]{@{}l@{}}BiLSTM CRF entity \\extraction recognition\end{tabular}           & 6.0\%  \\ \cline{4-5} 
       &      &  & \begin{tabular}[c]{@{}l@{}}Potential \\ relation marking\end{tabular}             & 0.9\%  \\ \cline{4-5} 
       &      &  & \begin{tabular}[c]{@{}l@{}}PCNN-ATT \\ relation extraction\end{tabular}           & 5.7\%  \\ 
\bottomrule
\end{tabular}%
\end{adjustbox}
\vspace{-2ex}
\end{table}

We measured a single-process procedure for all \cti reports.
The evaluation took 87.3 hours to finish, reaching a processing throughput of 24.7 reports per minute. 
With 11 articles added to the system every day, the expected daily workload is less than 30 seconds.
We show a performance breakdown analysis in ~\cref{tab:breakdown}. 
We notice that the extractors take most of the time and the dependency parsing is the bottleneck. 
A potential reason is that the sentence-wise dependency parsing for long content \cti report is time-consuming. 
As evidence, the dependency parsing for the source apt\_notes with an average content length of 32503 characters takes 22.0 seconds on average (88.5\% of total processing time for that source).
In contrast, for the source symantec\_vulnerability with an average content length of 332 characters, it takes 0.1 seconds on average (71.5\% of total processing time for that source).
These results show that \tool is efficient enough for real-world use cases.

\section{Related Work}
\label{sec:related}

\paragraph{\cti services and platforms}
Various platforms and services have been created to manage \cti. 
Platforms like AlienVault OTX \cite{alienvault-otx}, IBM X-Force \cite{ibmxforce}, PhishTank \cite{phishtank}, MISP \cite{misp}, and OpenCTI \cite{opencti} allow users to contribute, share, 
or manage \cti.
Unlike these platforms that require user contribution,
\tool gathers and aggregates threat knowledge automatically using AI-based techniques.
There are also platforms, such as OpenPhish \cite{openphish} and Abuse.ch \cite{abusech}, that collect threat knowledge automatically. 
However, they only focus on specific types of entities:
ThreatMiner focuses on low-level IOCs (\eg IPs, domains, and URIs). OpenPhish focuses on phishing URLs. Abuse.ch focuses on malware and botnets.
In addition, several studies have been proposed to better analyze \cti reports, such as understanding vulnerability reproducibility~\cite{mu2018understanding} and measuring threat knowledge quality (\eg consistency, accuracy, and coverage)~\cite{li2019reading,dong2019towards}.
Such research is orthogonal to \tool.

\paragraph{\cti formats and ontologies}
There exist open standard formats such as STIX~\cite{stix} and OpenIOC~\cite{openioc} for exchanging threat intelligence.
They are schemas rather than the large knowlede graph constructed by \tool that contains the actual knowledge.
The knowledge gathered by \tool can be easily converted into these formats for distribution.
MITRE ATT\&CK~\cite{mitre-attack} is a knowledge base for cyber adversary behaviors based on real-world observations.
It is manually curated by security experts and does not focus on automated knowledge extraction from \cti reports as done in \tool.
It also does not contain IOC relations.
There are some cyber ontologies~\cite{more2012knowledgebased, undercofer2003targetcentric, gregio2014ontology, syed2016uco} that support reasoning, but most of them only focus on sub-domains of threat knowledge, such as IDS \cite{undercofer2003targetcentric, more2012knowledgebased} and malware behavior \cite{gregio2014ontology}.
None of these works focus on automated threat knowledge extraction from natural language text.

\paragraph{Security information extraction}
Several works have been proposed to extract threat knowledge from text.
iACE~\cite{liao2016acing}
extracts IOCs 
from security articles using a graph mining technique.
ChainSmith~\cite{zhu2018chainsmith} further classifies the extracted IOCs into different attack campaign stages (\eg baiting, exploitation, installation, and C\&C) using neural networks.
TTPDrill \cite{husari2017ttpdrill} extracts threat actions from Symantec reports and maps them to pre-defined attack patterns.
EXTRACTOR~\cite{satvat2021extractor}, ThreatRaptor~\cite{gao2021enabling}, and HINTI~\cite{hinti} use various NLP techniques to extract IOC entities and IOC relations. 
These work focus extract only IOCs or IOC relations from a single \cti report.
In contrast, \tool extracts a wider range of entities (\eg threat actors, techniques, tools) and relations from multiple reports to construct a threat knowledge graph.

\section{Conclusion}

We presented \tool, a system for automated open-source cyber threat intelligence gathering and management. \tool automatically constructs a threat knowledge graph from \cti reports using AI-based techniques.
In future work, we aim to explore other types of security applications that can be enabled by \tool, such as intrusion detection and cyber threat hunting.

\paragraph{Acknowledgement}
This work was supported in part by the 2021 Cisco Research Award and the Commonwealth Cyber
Initiative (CCI).
Any opinions, findings, and conclusions made in this material are those of the authors and do not necessarily reflect the views of the funding agencies.

\bibliographystyle{ACM-Reference-Format}
% \balance
\bibliography{refs}

\newpage

% \section*{Appendix}

\begin{table*}[ht!]
\caption{List of \cti sources}
\label{table:oscti-sources-and-statistics}

\begin{adjustbox}{width=0.8\linewidth}

\begin{tabular}{lccl}
\toprule
\textbf{\cti Source} & \textbf{Number of Reports} & \textbf{Type} & \textbf{URL} \\ 
\midrule
apt_notes & 539 & APT Reports & \url{https://github.com/aptnotes/data} \\
attcybersecurity & 244 & Enterprise Security Blog & \url{https://cybersecurity.att.com} \\
ciscoumbrella & 478 & Enterprise Security Blog & \url{https://umbrella.cisco.com} \\
cloudflare & 1,791 & Enterprise Security Blog & \url{https://blog.cloudflare.com} \\
crowdstrike & 942 & Enterprise Security Blog & \url{https://www.crowdstrike.com} \\
csoonline & 1,258 & Enterprise Security Blog & \url{https://www.csoonline.com/} \\
darknet & 2,107 & Enterprise Security Blog & \url{https://www.darknet.org.uk} \\
fireeye & 209 & Enterprise Security Blog & \url{https://www.fireeye.com} \\
forcepoint & 1,190 & Enterprise Security Blog & \url{https://www.forcepoint.com} \\
hotforsecurity & 9,496 & Enterprise Security Blog & \url{https://hotforsecurity.bitdefender.com} \\
kasperskydaily & 3,350 & Enterprise Security Blog & \url{https://www.kaspersky.com} \\
krebsonsecurity & 2,129 & Personal Security Blog & \url{https://krebsonsecurity.com} \\
malwarebytes & 3,382 & Enterprise Security Blog & \url{https://blog.malwarebytes.com} \\
mcafee & 6,295 & Enterprise Security Blog & \url{https://www.mcafee.com} \\
nakedsecurity & 14,653 & Enterprise Security Blog & \url{https://nakedsecurity.sophos.com} \\
nccgroup & 520 & Enterprise Security Blog & \url{https://research.nccgroup.com} \\
paloalto & 3,284 & Enterprise Security Blog & \url{https://blog.paloaltonetworks.com} \\
recordedfuture & 1,537 & Enterprise Security Blog & \url{https://www.recordedfuture.com} \\
rsa & 71 & Enterprise Security Blog & \url{https://www.rsa.com} \\
securelist & 5,630 & Enterprise Security Blog & \url{https://securelist.com} \\
shneieronsecurity & 8,110 & Personal Security Blog & \url{https://www.schneier.com} \\
sophos & 1,822 & Enterprise Security Blog & \url{https://news.sophos.com} \\
spiderlabs & 1,401 & Enterprise Security Blog & \url{https://www.trustwave.com/en-us/resources/blogs/spiderlabs-blog/} \\
symantecthreatintelligence & 177 & Enterprise Security Blog & \url{https://symantec-enterprise-blogs.security.com/blogs/threat-intelligence/} \\
thehackernews & 8,432 & Enterprise Security Blog & \url{https://thehackernews.com} \\
threatpost & 5,427 & Enterprise Security Blog & \url{https://threatpost.com/} \\
trendmicro & 2,393 & Enterprise Security Blog & \url{https://blog.trendmicro.com} \\
trendmicrosecurityintelligence & 4,001 & Enterprise Security Blog & \url{https://blog.trendmicro.com/trendlabs-security-intelligence} \\
trustwave & 571 & Enterprise Security Blog & \url{https://www.trustwave.com/en-us/resources/blogs/trustwave-blog/} \\
unit42_paloalto & 645 & Enterprise Security Blog & \url{https://unit42.paloaltonetworks.com/} \\
webroot & 1,438 & Enterprise Security Blog & \url{https://www.webroot.com} \\
welivesecurity & 5,780 & Enterprise Security Blog & \url{https://www.welivesecurity.com} \\
zscaler & 770 & Enterprise Security Blog & \url{https://www.zscaler.com} \\
malwarebytes & 163 & Threat Encyclopedia & \url{https://blog.malwarebytes.com} \\
symantec_threats & 37,588 & Threat Encyclopedia & \url{http://asb-sngweb.symantec.com} \\
symantec_vulnerabilities & 7,431 & Threat Encyclopedia & \url{http://asb-sngweb.symantec.com} \\
kaspersky_threat & 1,430 & Threat Encyclopedia & \url{https://threats.kaspersky.com/en} \\
kaspersky_vulnerability & 1,968 & Threat Encyclopedia & \url{https://threats.kaspersky.com/en} \\
trendmicro_malware & 534 & Threat Encyclopedia & \url{https://www.trendmicro.com} \\
trendmicro_spam & 396 & Threat Encyclopedia & \url{https://www.trendmicro.com} \\
fsecure & 4,083 & Threat Encyclopedia & \url{https://www.f-secure.com} \\
\bottomrule
\end{tabular}
\end{adjustbox}
\end{table*}

\begin{table*}[t]
\centering
\caption{List of relation classes that \tool covers} 
\label{table:relation}
\begin{adjustbox}{width=1\linewidth}
\begin{tabular}{llll}
\toprule
\textbf{No} & \textbf{Relation Class} & \textbf{Explanation} & \textbf{Indicator Verbs} \\
\midrule
1           & use                                       & A bad actor, malware, or technique uses something to finish a goal. The action is general, without much detail. & use, take, utilize, employ                                                          \\
2           & execute                                   & An actor executes a specific tool, program, function, etc.                                                      & perform, parse, execute, conduct, run, calculate, carry out, call, initiate, launch \\
3           & enable                                    & A tool/technique enables one thing means this tool/technique makes this thing possible.                         & tunnel, allow, rely, provide, sign, attribute, harden,activate                      \\
4           & own                                       & One thing owns something means such thing contains something or is composed by something.                       & compose, include, consist, contain, inside                                          \\
5           & inject                                    & Typically “foo injects bar” means a bad actor foo inserts something malicious, bar into the targets.            & save, load, attack, install, write, embed, upload, inject, deploy, infect           \\
6           & alter                                     & “foo alters bar” means foo modifies or changes something bar on the targets to achieve some malicious goals.    & change, define, affect, compromise, change, configure, tamper, redirect             \\
7           & get                                       & The subject obtains some information, data, etc.                                                                & decrypt, retrieve, extract, download, obtain, send, receive, steal, access          \\
8           & keep                                      & To make something consistent by remaining or storing something.                                                 & persist, maintain, remain, store, host                                              \\
9           & spread                                    & To duplicate and send one thing (typically malware) from one place to other places.                             & spread, circulate, distribute, release, share, duplicate, propagate                 \\
10          & hide                                      & Foo hides bar means foo makes bar unseeable or undetectable.                                                    & encrypt, hide, obfuscate                                                            \\
11          & relate                                    & One thing has some relations or communications with another thing.                                              & {\color[HTML]{333333} link, match, relate, associate, communicate, connect, alias}  \\
12          & create                                    & To generate or make something that did not exist before.                                                        & compute, craft, create,build                                                        \\
13          & update                                    & Neutral word, typically means to update a program to a new version, or update the state.                        & modify, recreate, restructure                                                       \\
14          & break                                     & Foo breaks bar means foo stops or prevents bar                                                                  & delete, block, destroy, stop, circumvent, bypass, drop                              \\
15          & find                                      & To discover or locate the desired things from the whole dataset.                                                & select, find, search, detect, look for, scan                                        \\
16          & mitigate                                  & Typically means to alleviate bad influence.                                                                     & mitigate, resolve, protect                                                          \\
17          & aim                                       & A bad actor/malware aims at one thing means this thing is the target or victim.                                 & aim, target, attack, for                                                            \\                                                                             

\bottomrule
\end{tabular}
\end{adjustbox}

\end{table*}

\end{document}

%% file: commands-reduce.tex
\usepackage{color}
\usepackage{syntax}
\usepackage{amsmath,amsfonts}
\usepackage{booktabs}
\usepackage[T1]{fontenc}
\usepackage{xspace}
\usepackage{enumitem}

\usepackage{adjustbox}

\usepackage{mathptmx}

\usepackage[labelfont=bf,skip=5pt,figurename=Fig.]{caption}
\usepackage{subcaption}

\usepackage{hyperref} %change this later
\hypersetup{
	citebordercolor={0 1 0},
	citecolor=blue,
	colorlinks=false,
	filebordercolor={0 .5 .5},
	filecolor=green,
	linkbordercolor={1 0 0},
	linkcolor=blue,
	menubordercolor={1 0 0},
	pageanchor=true,
	pagebackref=true,
	pdfborder={0 0 1},
	pdfpagelabels=true,
	pdftex,
	plainpages=false,
	urlbordercolor={0 1 1},
	urlcolor=blue,
}

\usepackage[capitalize,nameinlink]{cleveref}
\hypersetup{%
	bookmarksnumbered, bookmarksopen=true, bookmarksopenlevel=1,%
}
\crefname{figure}{Fig.}{Figs.}
\crefname{listing}{Query}{Queries}
\crefname{section}{Section}{Sections}
\crefname{table}{Table}{Tables}
\crefname{BNF}{Grammar}{Grammars}
\crefname{algorithm}{Algorithm}{Algorithms}
\crefname{equation}{Equation}{Equations}

\usepackage{listings}

\lstdefinestyle{myStyleMain}{
    belowcaptionskip=1\baselineskip,
    breaklines=true,
    basicstyle=\footnotesize\ttfamily,
    keywordstyle=\bfseries,  % bold keywords
    backgroundcolor=\color{white},
    xleftmargin=8pt,
    morekeywords={MATCH, WHERE, RETURN, IN}  % add your keywords here
}

\usepackage{multirow}

\renewcommand{\paragraph}[1]{\smallskip\noindent\textbf{#1.}}

\newcommand{\tool}{\textsc{ThreatKG}\xspace}
\newcommand{\qatool}{\textsc{ThreatQA}\xspace}
\newcommand{\cti}{OSCTI\xspace}

\newcommand{\seckg}{TKG\xspace}

\newcommand{\tkr}{UTKR\xspace}
\newcommand{\tkrs}{UTKRs\xspace}
\newcommand{\eg}{e.g.,\xspace}
\newcommand{\ie}{i.e.,\xspace}

\newcommand{\ent}[1]{\textsf{#1}}